\documentclass[12pt]{article}

\begin{document}

\begin{center}

%{\title THERMAL RATIO OF THE DISORDER DEVIATION AND
% THE SPACE-TIME DECONFINED PHASE SIZE}
\noindent{\large THERMAL RATIO OF THE DISORDER DEVIATION AND
 THE SPACE-TIME DECONFINED PHASE SIZE}

\vskip 5mm

G.A. Kozlov

\vskip 5mm

{\small {\it
Bogoliubov Laboratory of Theoretical Physics\\
Joint Institute for Nuclear Research\\
141980 Dubna, Moscow region, Russia
}}

 {\it
E-mail: kozlov@thsun1.jinr.ru
}
\end{center}

\vskip 5mm

\begin{center}
\begin{minipage}{150mm}
\centerline{\bf Abstract}

A systematic study of the strongly interacting matter under extreme
conditions via the form of the thermal
ratio of the disorder deviation is presented.
The evolution of fermi- and bose- particles (quarks and gluons) is 
studied in the framework of multi-particle correlation and 
distribution functions to predict the size of the finite-temperature 
deconfined phase.\\


\end{minipage}
\end{center}

\vskip 10mm

\section{Introduction}
\ \ \ \ \  The existence of a deconfined phase of gluons and
quarks has been predicted by quantum
 chromodynamics (QCD) [1]. Aspects of QCD at finite density
 are very important as well as
instructive in the sense of the observation of quark-gluon
plasma (QGP), especially from the heavy vector meson (HVM) suppression,
strangeness enhancement, hadron yield distributions versus
temperature and dilepton excess. There is large amount of consistency
among the different signatures of deconfined phase transition to
a new excited state of matter.
In the case of HVM suppression (e.g., $c\bar{c}$ bound state
suppression), the CERN-SPS experiments clearly show that it's
visible in Pb-Pb collisions starting from a definite value of
the energy density in the right range. Enhanced strange
particle production in Pb-Pb collisions offers another
indication for the phase transition: in the deconfined phase
strange quarks are more easily produced. The threshold for the
producing strange quarks is much lower than that for strange
hadrons. In addition, the mass of the strange quark goes down in
the case of the restoration of the chiral symmetry. Hence, the
observation of large ratios for strange particles is considered as
a remnant of the unconfined phase. The dilepton abundance, while
well explained by ordinary soucers in all other cases, for Pb-Pb
collisions a neat excess is instead observed.

There is a very popular point of view in literature that a deconfinement
phase transition is predicted to occur at the typical energy scale involved,
the temperature $T_{c}\sim$ QCD scale $\Lambda_{QCD}
\sim$ the pion mass  $m_{\pi}\sim$ the mass of a strange quark $m_{s}\sim$
140-200 MeV.
Putting all indications above mentioned together, the case of a
high density state of matter onsetting at a critical temperature
around 170 MeV is reasonably suitable.

 Among the issues related to QGP, we attract attention to the
problem of deconfined phase through the calculation of correlation and
distribution functions [2] in the thermal theory of quantized fields.
 We consider the semiphenomenological model for the
deconfinement existence within the framework of the Langevine-type equation.
 To do this, we follow the standard theory of quantized fields replacing:\\
1. the asymptotic field operators and\\
2. the vacuum expectation values\\
by\\
1. the thermal field operators and\\
2. the thermal statistical averages,\\
respectively, in order to formulate correlation and distribution functions of
produced particles.

  The method of
Langevin equation and its extensions to the quantal case have been
suggested and considered in papers [3-5] and [6-8], respectively.
 We propose that real physical processes to happen
in the finite-temperature QCD should be replaced by a one-constituent
(e.g., gluon or quark) propagation provided by a special
kernel operator (in the master evolution equation) to be considered as an
input of the model and disturbed by the random force $F$.
We assume $F$ to be the external source  proposed as
both a c-number function and an operator. The master equation is an
operator one, so that there appear new additional issues about the
commutation relations and the ordering of operators, which do not exist in
the classical case [8].

   Based on the thermal operator-field technique, we introduce a
thermal ratio of the disorder deviation (TRDD),
reflecting the degree of deviation, from unity, of the ratio of the
two-particle
thermal momentum-dependent distribution to two one-particle thermal
distribution functions of produced particles, gluons and/or
quarks (g/q) in a partly deconfined phase state. We study the
four-momentum correlations of identical particles which can be both
useful and instructive to infer the shape of the
particle emitter-source. We estimate the sensitivity of the TRDD-functions to
the size of the emitter.
Within these features, the canonical formalism in a
stationary state in the thermal equilibrium (SSTE) is formulated, and a closed
structural resemblance between the SSTE and standard quantum field theory
is revealed.\\
  To clarify the internal structure of the disordering of particles,
 we use the consistent approach
based on the evolution of dynamical variables as well as the extension to
different modes provided by virtual transitions.

\section{ Distribution functions. Evolution equation.}

   Let us consider a hypothetical system of the quark-gluon excited local
thermal phase in QCD where a canonical operator $a$ and
its Hermitian conjugate $a^{+}$  occur.
We formulate the
distribution functions (DF) of produced particles (gluons and quarks) in
terms of
point-to-point equal time temperature-dependent thermal
correlation functions (CF) of two operators
%\begin{eqnarray}
%\label{e1}
$$w(\vec{k},\vec{k}^{\prime},t;T)=\langle a^{+}(\vec{k},t
)\,a(\vec{k}^{\prime},t)\rangle = $$
$$= Tr [a^{+}(\vec{k},t)\,a(\vec{k}^{\prime},t)e^{-H\beta}]/Tr (e^{-H\beta})\ .$$
%\end{eqnarray}
 Here, $\langle ...\rangle$ means the procedure of thermal statistical
averaging; $\vec{k}$ and $t$ are, respectively, momentum and time variables,
$e^{-H\beta}/Tr(e^{-H\beta})$ stands for the standard density operator in the
equilibrium and the Hamiltonian $H$ is given by the squared form of the
annihilation $a_{p}$ and creation $a_{p}^{+}$ operators for bose- and fermi-
particles,
$H=\sum_{p}\epsilon_{p}a^{+}_{p}a_{p}$ (the energy $\epsilon_{p}$ and
operators $a_{p}$, $a_{p}^{+}$ carry some index $p$ [9], where
$p_{\alpha}=2\pi\ n_{\alpha}/L, n_{\alpha}=0,\pm 1,\pm 2, ...; V=L^{3}$ is the
volume of the system considered); $\beta$
is the inverse temperature of the environment, $\beta=1/T$.
     The standard canonical commutation relation
%\begin{eqnarray}
%\label{e1}
$${\left\lbrack a(\vec{k},t),a^{+}(\vec{k}^{\prime},t)
\right\rbrack}_{\pm}=\delta^{3}(\vec{k}-\vec{k}^{\prime})$$
%\end{eqnarray}
at every time t is used as usual for Bose (-) and Fermi (+)-operators.

 The probability
to find two particles, gluons or quarks, with momenta $\vec{k}$ and $\vec{k}^
{\prime}$ in the same event at the time $t$ normalized to the single
spectrum of these particles is:
%\begin{eqnarray}
%\label{e2}
$$R(\vec{k},\vec{k}^{\prime},t)=W(\vec{k},\vec{k}^{\prime},t)/[W(\vec{k},t)
\cdot W(\vec{k}^{\prime},t)]\ . $$
%\end{eqnarray}
 Here, the one-particle thermal DF is defined as
%\begin{eqnarray}
%\label{e3}
$$W(\vec{k},t)=\langle b^{+}(\vec{k},t)\ b(\vec{k},t)\rangle\ ,
$$
%\end{eqnarray}
where
%\begin{eqnarray}
%\label{e4}
$$b(\vec{k},t)=a(\vec{k},t)+\phi(\vec{k},t) $$
%\end{eqnarray}
 under an assumption of
occurrence of the random source-function $\phi(\vec{k},t)$ being an operator,
in general. The two-particle DF $W(\vec{k},\vec{k}^{\prime},t)$ looks like
%\begin{eqnarray}
%\label{e5}
$$W(\vec{k},\vec{k}^{\prime},t)=\langle b^{+}(\vec{k},t)\ b^{+}(\vec{k}
^\prime,t)\ b(\vec{k},t)\ b(\vec{k}^{\prime},t)\rangle\ . $$
%\end{eqnarray}
 The evolution
properties of propagating particles in a randomly distributed
environment comes from the evolution equations
\begin{eqnarray}
\label{e6}
i\ \partial_{t}b(\vec{k},t)+A(\vec{k},t)=F(\vec{k},t)+P\ ,
\end{eqnarray}
\begin{eqnarray}
\label{e7}
-i\ \partial_{t}b^{+}(\vec{k},t)+A^{*}(\vec{k},t)=F^{+}(\vec{k},t)+P\
,
\end{eqnarray}
where both $b$ and $b^{+}$ are the special mode operators of
the quark and gluon fields [10], $P$ and $F(\vec{k},t)$
stand for the stationary external force and the random one, respectively,
both acting
from the environment. The only operator $F$ has a zeroth value of the
statistical average, $\langle{F}\rangle=0$. The interaction of the particles
considered with the surrounding ones as well as providing the propagation
is given by the operator $A(\vec{k},t)$ which can be defined as the one
closely related to the dissipation force:
\begin{eqnarray}
\label{e8}
A(\vec{k},t)=\int_{-\infty}^{+\infty}K(\vec{k},t-\tau)\ b(\vec{k},\tau)\ d\tau\ .
\end{eqnarray}
 An interplay of quarks and gluons with
surrounding particles is embedded into the interaction complex kernel
$K(\vec{k},t)$, while the real physical transitions are provided by the
random source operator $F(\vec{k},t)$ ( see eqns. (\ref{e6}) and  (\ref{e7})).
The random evolution field operator $K(\vec{k},t)$ in (\ref{e8}) stands for
the random noise and it is assumed to vary stochastically with a $\delta$-
like equal time correlation function [10]
%\begin{eqnarray}
%\label{e9}
$$\langle K^{+}(\vec{k},\tau)\ K(\vec{k}^\prime,\tau)\rangle =2\
{(\pi\rho)}^{1/2}\ \kappa\ \delta (\vec{k}-\vec{k}^\prime)\ ,
$$
%\end{eqnarray}
where both the strength of the noise $\kappa$ and the positive
constant $\rho\rightarrow\infty$ define the effect of the
Gaussian noise on the evolution of quarks and gluons in the
thermalized environment.

  The formal solution of (\ref{e6}) in the operator form
in $S(\Re_{4})$ ($k^{\mu}=(\omega=k^0,k_{j}))$ is
%\begin{eqnarray}
%\label{e10}
$$\tilde b(k_{\mu})=\tilde a(k_{\mu})+\tilde\phi(k_{\mu})\ , $$
%\end{eqnarray}
where the operator $\tilde a(k_{\mu})$ is expressed via the Fourier
transformed operator $\tilde F(k_{\mu})$ and the Fourier transformed kernel
function $\tilde K(k_{\mu})$ (coming from (\ref{e8})) as
%\begin{eqnarray}
%\label{e12}
$$\tilde a(k_{\mu})=\tilde F(k_{\mu})\cdot [\tilde K(k_{\mu})-\omega]^{-1}\
,$$
%\end{eqnarray}
while the function $\tilde\phi(k_{\mu})$ $\sim P\cdot [\tilde K(k_{\mu})-\omega]^{-1}$.
 The random
force operator $F(\vec{k},t)$ can be expanded by using the Fourier integral
\begin{eqnarray}
\label{e12}
F(\vec{k},t)=\int_{-\infty}^{+\infty}\frac{d\omega}{2\pi}\ \psi(k_{\mu})\
\hat c(k_{\mu})\ e^{-i\omega t}\ ,
\end{eqnarray}
where the form $\psi(k_{\mu})\cdot\hat c(k_{\mu})$ is just the Fourier
operator $\tilde F(k_{\mu})=\psi(k_{\mu})\cdot\hat c(k_{\mu})$ and the
canonical operator $\hat c(k_{\mu})$ obeys the commutation relation
%\begin{eqnarray}
%\label{e21}
$${\left\lbrack\hat c(k_{\mu}), \hat c^{+}(k_{\mu}^{\prime})\right\rbrack}_{\pm}=
\delta^{4}(k_{\mu}-k^{\prime}_{\mu})\ .$$
%\end{eqnarray}
The function $\psi(k_{\mu})$ in (\ref{e12}) is determined by the condition [10]
$$\int_{-\infty}^{+\infty}\frac{d\omega}{2\pi}
{\left [\frac{\psi(k_{\mu})}{\tilde K(k_{\mu})-\omega}\right ]}^{2}=1\ . $$

\section{TRDD and the space-time size.}
  The enhanced probability for emission of two identical
particles is given by the ratio $R$ of DF in $S(\Re_{4})$ as follows:
\begin{eqnarray}
\label{e13}
R_{b/f}(k_{\mu},k_{\mu}^\prime;T)=\frac{\tilde W(k_{\mu},k_{\mu}^\prime;T)}
{\tilde W(k_{\mu})\cdot\tilde W(k_{\mu}^\prime)}\ ,
\end{eqnarray}
where $\tilde W(k_{\mu},k_{\mu}^\prime;T)=\langle\tilde b^{+}(k_{\mu})\ \tilde b^
{+}(k_{\mu}^\prime)\ \tilde b(k_{\mu})\ \tilde b(k_{\mu}^\prime)\rangle
$ and $\tilde W(k_{\mu})=\langle\tilde b^{+}(k_{\mu})\ \tilde b(k_{\mu})\rangle
$.
 Using the Fourier solution of equation (\ref{e6}) in $S(\Re_{
4})$, one can get R-ratios for DF obeying to Bose-
\begin{eqnarray}
\label{e14}
R_{b}(k_{\mu},k_{\mu}^\prime;T)=1+D_{b}(k_{\mu},k_{\mu}^\prime;T)
\end{eqnarray}
and Fermi-particles
\begin{eqnarray}
\label{e15}
R_{f}(k_{\mu},k_{\mu}^\prime;T)=R_{b}(k_{\mu},k_{\mu}^\prime;T)-
2\ \frac{\Xi(k_{\mu})\cdot\Xi(k_{\mu}^\prime)}{\tilde W(k_{\mu})
\cdot\tilde W(k_{\mu}^\prime)}\ ,
\end{eqnarray}
where
\begin{eqnarray}
\label{e16}
D_{b}(k_{\mu},k_{\mu}^\prime)=\frac{\Xi(k_{\mu},k_{\mu}^\prime)[\Xi(k_{\mu}^
\prime,k_{\mu})+\tilde\phi^{+}(k_{\mu}^\prime)\tilde\phi(k_{\mu})]+
\Xi(k_{\mu}
^\prime,k_{\mu})\tilde\phi^{+}(k_{\mu})\tilde\phi(k_{\mu}^\prime)}{
\tilde W(k_{\mu})\cdot\tilde W(k_{\mu}^\prime)}
\end{eqnarray}
and the two-particle CF $\Xi(k_{\mu},k_{\mu}^\prime)$ looks like
\begin{eqnarray}
\label{e17}
\Xi(k_{\mu},k_{\mu}^\prime)=\langle\tilde a^{+}(k_{\mu})\ \tilde a(k_{\mu}^
\prime)\rangle \cr
=\frac{\psi^{*}(k_{\mu})\cdot\psi(k_{\mu}^\prime)}{[\tilde K^{*}(k_{\mu})-
\omega]\cdot[\tilde K(k_{\mu}^\prime)-\omega^\prime]}\cdot\langle\hat c^{+}
(k_{\mu})\ \hat c(k_{\mu}^\prime)\rangle\ .
\end{eqnarray}
 Using the  Kubo-Martin-Schwinger condition ( $\mu$ is the chemical potential)
%\begin{eqnarray}
%\label{e18}
$$\langle a(\vec{k}^\prime,t^\prime)\ a^{+}(\vec{k},t)\rangle = \langle a^{+}
(\vec{k},t)\ a(\vec{k}^\prime,t-i\beta)\rangle\cdot \exp(-\beta\ \mu)\
,$$
%\end{eqnarray}
the thermal statistical averages for the $\hat c(k_{\mu})$-operator
should be presented in the following form:
%\begin{eqnarray}
%\label{e19}
$$\langle\hat c^{+}(k_{\mu})\ \hat c(k_{\mu}^\prime)\rangle =\delta^{4}(k_{\mu}-
k_{\mu}^\prime)\cdot n(\omega,T)\ , $$
%\end{eqnarray}
%\begin{eqnarray}
%\label{e20}
$$\langle\hat c(k_{\mu})\ \hat c^{+}(k_{\mu}^\prime)\rangle =\delta^{4}(k_{\mu}-
k_{\mu}^\prime)\cdot [1\pm n(\omega,T)]$$
%\end{eqnarray}
for Bose (+)- and Fermi (-)-statistics, where $n(\omega,T)=\{\exp[(\omega-\mu)
\beta]\pm 1\}^{-1}$. Inserting CF (\ref{e17}) into
(\ref{e16}) and taking into account that the
$\delta^{4}(k_{\mu}-k_{\mu}^\prime)$-function should be changed by the smooth
sharp
function $\Omega(r)\cdot \exp(-q^2/2)$, one can get the following expression
for the $D_{b}$-function
\begin{eqnarray}
\label{e21}
D_{b}(k_{\mu},k_{\mu}^\prime;T)= \lambda(k_
{\mu},k_{\mu}^\prime;T)\ \exp(-q^2/2) \cr
\times [n(\bar{\omega},T)\Omega(r) \exp(
-q^2/2)+\tilde\phi^{*}(k_{\mu}^\prime)\tilde\phi(k_{\mu})+\tilde\phi^{*}(k_{
\mu})\tilde\phi(k_{\mu}^\prime)]\ ,
\end{eqnarray}
where
%\begin{eqnarray}
%\label{e37}
$$\lambda(k_{\mu},k_{\mu}^\prime;T)=\frac{\Omega(r)}{\tilde W(k_{\mu})\cdot
\tilde W(k_{\mu}^\prime)}\cdot n(\bar{\omega},T)\ ,\ \bar{\omega}=\frac{1}{2}
(\omega+\omega^\prime)\ .$$
%\end{eqnarray}
The function $\Omega(r)\cdot n(\omega;T)\cdot \exp(-q^2/2)$ describes the
space-time size of the QGP fire-ball. Choosing the z-axis along the
two-heavy-ion collision axis one can put
$$q^2=(r_{0}\cdot Q_{0})^2+(r_{z}\cdot Q_{z})^2+
(r_{t}\cdot Q_{t})^2\ , $$
$$Q_{\mu}=(k-k^\prime)_{\mu}, Q_{0}=\epsilon_
{\vec{k}}-\epsilon_{\vec{k}^\prime}, Q_{z}=k_{z}-k_{z}^\prime, Q_{t}=
{{[(k_{x}-k_{x}^\prime)}^2+{(k_{y}-k_{y}^\prime)}^2]}^{1/2}\ , $$
$$\Omega(r)\sim r_{0}\cdot r_{z}\cdot r_{t}^2\ ,$$
where $r_{0}$, $r_{z}$ and $r_{t}$ are time-like, longitudinal and transverse
"size" components of the QGP fire-ball.
Formally, the function $D_{b}$ (\ref{e21}) is the positive one ranging from
0 to 1.
The quantitative information (longitudinal $r_{z}$ and transverse $r_{t}$
components of the QGP spherical volume, the temperature T of the environment)
could be extracted by fitting the theoretical formula (\ref{e21}) to
the measured TRDD function and estimating the errors of the fit parameters.
Formula (\ref{e21}) indicates that a chaotic g/q source emanating from the
thermalized g/q fireball exists. Hence, the measurement of the space-time
evolution of the g/q source would provide information of the g/q emission
process and the general reaction mechanism. In formula (\ref{e21}) for the
$D_{b}$- function, the temperature of the environment enters through the
two-particle CF $\Xi(k_{\mu},k_{\mu}^\prime;T)$. If T is unstable the
$R_{b/f}$-functions (\ref{e13}) will change due to a change of
DF $\tilde W$ which, in fact, can be considered as an effective density
of the g/q source.
   Formula (\ref{e14}) looks like the following expresion for the
experimental R-ratio using a source parametrization:
%\begin{eqnarray}
%\label{e22}
$$R_{T}(r)=1+\lambda_{T}(r)\cdot \exp(-{r_{t}^2\cdot Q_{t}^2}/2 -{r_{z}^2
\cdot Q_{z}^2}/2)\ , $$
%\end{eqnarray}
where $r_{t}(r_{z})$ is the transverse (longitudinal) radius parameter of the
source with respect to the beam axis, $\lambda_{T}$ stands for the effective
intercept parameter (chaoticity parameter) which has a general dependence of
the mean momentum of the observed particle pair. Here, the dependence on the
source lifetime is omitted. Since $0<\lambda_{T}<1$, one can conclude that the
effective function $\lambda_{T}$ can be interpreted as a function of the
core particles to all particles produced. The chaocity parameter
$\lambda_{T}$ is the temperature-dependent and the positive one defined by
%\begin{eqnarray}
%\label{e23}
$$\lambda_{T}(r)=\frac{{\vert\Omega(r)\
n(\bar{\omega};T)\vert }^2}{\tilde W(k_{\mu})\cdot\tilde W(k_{\mu}^\prime)
}\ .$$
%\end{eqnarray}
%where $\tilde W(k_{\mu})\cdot\tilde W(k_{\mu}^\prime)$ is replaced by
%$\tilde W_{0}(k_{\mu},k_{\mu}^\prime)$ for convenience regarding the
%point of view that one can distinguish different particles.

   Comparing (\ref{e16}) and (\ref{e21}) one can identify
%\begin{eqnarray}
%\label{e40}
$$\Xi(k_{\mu},k_{\mu}^\prime)=\Omega(r)\cdot n(\bar{\omega};T)
\cdot \exp(-q^2/2)\ .$$
%\end{eqnarray}
There is no a satisfactory tool to derive the precise analytical
form of the random source function $\tilde\phi(k_{\mu})$ in
(\ref{e16}), but one can put (see (\ref{e17}) and taking into
account $\tilde\phi(k_{\mu})$ $\sim P\cdot [\tilde
K(k_{\mu})-\omega]^{-1}$) [11,10]
%\begin{eqnarray}
%\label{e24}
$$\tilde\phi(k_{\mu})={[\alpha\cdot \Xi(k_{\mu})]}^{1/2}\ , $$
%\end{eqnarray}
where $\alpha$ is of the order $O\left
(P^2/n(\omega,T)\cdot{\vert\psi(k_{\mu})\vert}^2\right )$.
 Thus,
\begin{eqnarray}
\label{e25}
D_{b}(q^2;T)=\frac{\tilde\lambda^{1/2}(\bar{\omega};T)}{(1+\alpha)(1+
\alpha^\prime)}e^{-q^{2}/2}\left [\tilde\lambda^{1/2}(\bar{\omega};T)
e^{-q^{2}/2}+2{(\alpha\alpha^\prime)}^{1/2}\right]\ .
\end{eqnarray}
It is easily to see that in the vicinity of $q^{2}\approx 0$ one can get
the full correlation if $\alpha =\alpha^\prime =0$ and $\tilde\lambda (\bar
{\omega};T)$=1. Putting $\alpha =\alpha^\prime$ in (\ref{e25}) we find
the formal lower
bound on the space-time dimensionless size of the fire-ball for Bose-system:
%\begin{eqnarray}
%\label{e45}
$$q^2_{b}\geq\ln\frac{\tilde\lambda (\bar{\omega};T)}{{[\sqrt{(\alpha+1)^{2}+
\alpha^2}-\alpha]}^2}\ .$$
%\end{eqnarray}
In the case of Fermi-particles, the following restriction on $q^{2}_{f}$ is
valid (see (\ref{e15}))
%\begin{eqnarray}
%\label{e46}
$$\ln\frac{\tilde\lambda (\bar{\omega};T)}{{[\sqrt{2\alpha(\alpha+1)+3}
-\alpha]}^2}\ \leq q^2_{f}\ \leq\ln\frac{\tilde\lambda (\bar{\omega};T)}
{{[\sqrt{\alpha^2+2}-\alpha]}^2}\ .$$
 In fact, the function $D_{b}(k_{\mu},k_{\mu}^\prime;T)$ in (\ref{e21})
could not be observed because of some
model uncertainties. In the standard consideration, the TRDD-function has
to contain
a background contribution as well as other physical particles (resonances)
which have not been included in the calculation of the $D_{b}$- function.
In order to be close to the experimental data, one has to expand the $D_{b}$-
function as projected on some well-defined function (in $S(\Re_{4})$)
of the relative momentum of two particles produced in heavy-ion collisions
$D_{b}(k_{\mu},k_{\mu}^\prime;T)\rightarrow D_{b}(Q_{\mu}^{2};T).$
Thus, it will be very instructive to use the polynomial expansion which is
suitable to avoid any uncertainties as well as characterize the degree of
deviation from the Gaussian distribution, for example. In $(-\infty,+\infty)$,
a complete orthogonal set of functions can be obtained with the help of the
Hermite polynomials in the Hilbert space of the square integrable functions
with the measure $d\mu(z)=\exp(-z^2/2)dz$. The function $D_{b}$ corresponds to
this class if
$$\int_{-\infty}^{+\infty}dq \exp(-q^2/2)\ \vert D_{b}(q)\vert^n\ <\infty\ ,
 n=0,1,2,...\ .$$
The expansion in terms of the Hermite polynomials $H_{n}(q)$
\begin{eqnarray}
\label{e26}
D_{b}(q)=\lambda\sum_{n}\ c_{n}\cdot H_{n}(q)\cdot \exp(-q^2/2)\
\end{eqnarray}
is well suited for the study of possible deviation from both the experimental
shape and the exact theoretical form of the TRDD function $D_{b}$
(\ref{e21}). The coefficients $c_{n}$ in (\ref{e26}) are defined via
the integrals over
the expanded functions $D_{b}$ because of the orthogonality condition
$$\int_{-\infty}^{+\infty} H_{n}(x)\ H_{m}(x)\ \exp(-x^2/2)\ dx=\delta_{n,m}
\ .$$
Thus, the observation of the two-particle correlation
(both for Bose- and Fermi-symmetrization) enable to extract the
properties of the structure of
$q^2$, i.e. the space-time size of QGP formation.

In order to be close to an experiment one has to replace
$R_{b,f}$ functions (\ref{e14}), (\ref{e15}) with respect to the
cylindrical symmetry angles $\theta$ and $\phi$ which are
non-observable ones at fixed $Q_{t}$:
$$
R_{b,f}(k_{\mu},k_{\mu}^{\prime};T)\rightarrow\bar{R}_{b,f}(Q_{t};T)=
N^{-1}\int dq_{t}\,dQ_{z}\,d\theta\, d\phi\,\tilde
{W}(k_{\mu},k_{\mu}^{\prime};T)\ ,$$
where
$$ N=\int dq_{t}\,dQ_{z}\,d\theta\, d\phi\,\tilde
{W}(k_{\mu})\,\tilde{W}(k_{\mu}^{\prime})\, ,$$
$$q_{t}=\frac{1}{\cos\theta+\sin\theta}
\,\left\{k_{x}+k_{y}\mp\frac{1}{2}Q_{t}\left [\cos (\theta+\phi)+\sin
(\theta+\phi)\right ]\right\}\ .$$
Then, $\bar{R}_{b,f}(Q_{t};T)=1+\bar{D}_{b,f}(Q_{t};T)$ with
$$\bar{D}_{b}(Q_{t};T)=\frac{\bar{N}^{-1}(T)}{(1+\alpha)(1+\alpha^{\prime})}\,
\exp [-(r_{t}Q_{t})^2]\,F_{b}(Q_{t};T)\, ,$$
$$\bar{D}_{f}(Q_{t};T)=\frac{1}{(1+\alpha)(1+\alpha^{\prime})}\,\left\{\bar{N}^{-1}(T)
\exp [-(r_{t}Q_{t})^2]\,F_{b}(Q_{t};T)-2\right\}\, ,$$
$$F_{b}(Q_{t};T)=\int dq_{t}\,dQ_{z}\,d\theta\,
d\phi\,n^{2}(\bar{\omega};T)\,e^{-\beta_{0Z}}[1+2\sqrt{\alpha\alpha^{\prime}
\tilde{\lambda}^{-1}(\bar{\omega};T)}\,e^{q^{2}/2}]\, ,$$
$$\beta_{0Z}\equiv (r_{0}Q_{0})^2+(r_{z}Q_{z})^2\, ,$$
$$\bar{N}(T)=\int dq_{t}\,dQ_{z}\,d\theta\,
d\phi\,n(\omega;T)\,n(\omega^{\prime};T)\,.$$
To avoid the trivial result, one should restrict the
$\alpha$-parameter as
$$\alpha\geq\tilde{\lambda}^{-1/2}(\bar{\omega};T)\,e^{q^{2}/2}-\frac{1}{2}
\tilde{\lambda}^{1/2}(\bar{\omega};T)\,e^{-q^{2}/2}\, .$$
In conclusion of this section, the experimental data are quite
desired to make the analysis of the particle fluctuation via
TRDD.

\section{Conclusion}
1. We investigated the finite temperature DF (of produced
identical particles, gluons and quarks) which can be both useful and
instructive to infer the shape
of the gluon/quark source-emitter. In fact, we have presented the method
of extracting the intercept and source parameters from the shape of the
TRDD-function.\\
2. The relations between the CF $\Xi(k_{\mu},k_{\mu}^\prime)$ (\ref{e17})
and the full
$R$-functions for Bose (\ref{e14})- and Fermi (\ref{e15})-particles at the
stage of the freeze-out are obtained. We have shown the sensitivity of the
correlation functions to the space-time geometry of the source-emitter (\ref
{e21}). The TRDD- function $D_{b}$ describes the size and shape
of the space-time domain where the secondary observed particles are
generated.\\
3. One can conclude that formally, the deconfined phase size scale can be
 determined by the
evolution behavior of the field operators and the critical temperature
$T=T_{c}$ (see formula (\ref{e21})).\\
4. Since, the TRDD-function $D_{b}$ is the positive
one and restricted by 1, we expect that the $R$-ratio at too small values of
$Q_{\mu}$ starts from the fixed point $R(Q_{\mu}\rightarrow 0)=2-\epsilon$
$(\epsilon\rightarrow +0)$ and then falls down (with the Gaussian shape) up
to unity over some momentum scale interval of an order of the inverse source
size.\\

\end{document}